\documentclass[submission,copyright,creativecommons]{eptcs}
\providecommand{\event}{PrePost 2016} % Name of the event you are submitting to
\usepackage{breakurl}             % Not needed if you use pdflatex only.

\usepackage{amssymb}
\usepackage{graphicx}
\usepackage{url}
\usepackage{booktabs}
\usepackage[space]{grffile}
\usepackage{color}
\definecolor{Light}{gray}{.95}
\usepackage[table]{colortbl}
\usepackage{amsmath}
\usepackage{yhmath}
\def\halfthinspace{\relax\ifmmode\mskip.5\thinmuskip\relax\else\kern.8888em\fi}

\newcommand {\absoracle}{\Omega}
\newcommand {\reloracle}{\omega}
\newcommand {\defin}{\delta}

\newcommand {\emptyrelation}{\phi}

\newcommand {\tabeq}{\hspace*{0.2in}}

\newcommand {\nln}{@{}l@{}}
\newlength{\interligne}
\newcommand {\dpreuve}{\dimen123=\linewidth \dimen124=\linewidth
\advance\dimen123 by -20mm \advance\dimen124 by -5mm
\advance\dimen123 by -\mathindent \advance\dimen124 by -\mathindent
\setlength{\interligne}{\baselineskip}
\setlength{\baselineskip}{1.2\baselineskip}
    \begin{tabbing} 
    \hspace*{\mathindent}\= \hspace*{5mm}\= \kill 
    \+ \kill}
\newcommand {\fpreuve}{\end{tabbing}
    \setlength{\baselineskip}{\interligne}}

\newcommand {\dpreuveitem}{\begin{tabbing} 
    \hspace*{5mm}\= \hspace*{15mm}\= \kill}
\newcommand {\fpreuveitem}{\end{tabbing}}

\newcommand {\dspecitem}{\begin{tabbing} 
    \hspace*{5mm}\=\hspace*{5mm}\=\hspace*{5mm}\=
    \hspace*{5mm}\=\hspace*{5mm} \kill}
\newcommand {\fspecitem}{\end{tabbing}}

\def\[{\relax\ifmmode\@badmath\else\begin{trivlist}\item[]\leavevmode
  \hbox to\linewidth\bgroup$ \displaystyle
  \hskip\mathindent\bgroup\fi}
\def\]{\relax\ifmmode \egroup $\hfil \egroup \end{trivlist}\else \@badmath \fi}

\newtheorem {proposition}{Proposition}
\newtheorem {definition}{Definition}

\newcommand {\refines}{\sqsupseteq}

\newcommand {\dom}{\mbox{\it dom}}

\newcommand {\dtabin}{\begin{tabbing} 
    \hspace*{\mathindent}\= \kill \+ \kill}
\newcommand {\ftabin}{\end{tabbing}}
\newcommand {\dspec}{\begin{tabbing} 
    \hspace*{\mathindent}\= \hspace*{5mm}\=\hspace*{5mm}\=\hspace*{5mm}\=
                            \hspace*{5mm}\=\hspace*{5mm} \kill 
    \+ \kill}

\newcommand {\fspec}{\end{tabbing}}

\newcommand {\modk}{~{\bf mod}~}

\newlength {\longueurtop}
\newcommand {\initlongueurtop}{\setlength{\longueurtop}{\topsep}}
\newcommand {\topzero}{\setlength{\topsep}{0pt}}
\newcommand {\topdefaut}{\setlength{\topsep}{\longueurtop}}
\newcommand {\debuttab}{ \initlongueurtop \topzero \begin{tabbing} }
\newcommand {\fintab}{ \end{tabbing} \topdefaut }

\title{Program Repair by Stepwise Correctness Enhancement}
\author{Nafi Diallo
\institute{CCS, NJIT, Newark NJ USA}
\email{ncd8@njit.edu}
\and
Wided Ghardallou
\institute{FST, UTM, Tunis, Tunisia}
\email{wided.ghardallou@gmail.com}
\and
Ali Mili
\institute{CCS, NJIT, Newark, NJ USA}
\email{mili@njit.edu}
}
\def\titlerunning{Program Repair by Correctness Enhancement}
\def\authorrunning{N. Diallo, W. Ghardallou, A. Mili}
\begin{document}
\maketitle
%\mainmatter  % start of an individual contribution
%\thanks{Acknowledgement:
%This publication was made possible by a grant from the Qatar National
%Research Fund, NPRP04-1109-1-174.  Its contents are solely the responsibility of 
%the authors and do not necessarily represent the official views of the QNRF.}}

\begin{abstract}
Relative correctness is the property of a program to be more-correct than
another with respect to a given specification.  Whereas the traditional
definition of (absolute) correctness divides candidate program into two
classes (correct, and incorrect), relative correctness arranges candidate
programs on the richer structure of a partial ordering.  In other venues
we discuss the impact of relative correctness on program derivation, and
on program verification.  In this paper, we discuss the impact of relative
correctness on program testing; specifically, we argue that when we remove
a fault from a program, we ought to test the new program for relative 
correctness over the old program, rather than for absolute correctness.
We present analytical arguments to support our position, as well as
an empirical argument in the form of a small program whose faults are
removed in a stepwise manner as its relative correctness rises with
each fault removal until we obtain a correct program.
\end{abstract}

\subsection*{Keywords}
Program correctness,
Relative correctness,
Absolute correctness,
Program repair.

\section{Relative Correctness and Quality Assurance Methods}

Relative correctness is the property of a program to be more-correct
than another with respect to a given specification.  Intuitively,
$P'$ is more-correct than $P$ with respect to 
specification $R$ if and only if $P'$ obeys
$R$ more often (for a larger set of inputs) than $P$, and violates $R$
less egregiously (in fewer ways)
than $P$.  

Traditionally, we distinguish between two categories of candidate programs
for a given specification:  correct programs, and incorrect programs; but
the introduction of relative correctness enables us to generalize this
binary classification into a richer structure
that ranks all candidate programs by means of a
partial ordering whose maximal elements are (absolutely) correct.

Also, in our quest for enhancing program quality, we have
traditionally
used static analysis methods and dynamic testing methods for distinct
purposes: 
\begin{itemize}
\item Program verification methods are applied to correct
programs to ascertain their correctness; they are of little use
when applied to incorrect programs, because even when a
proof fails, we cannot conclude that the program is incorrect
(the proof may have failed because the documentation of the
program in terms of intermediate assertions and invariant
assertions is inadequate).
\item Program testing methods 
are applied to incorrect programs to expose their faults and
remove them; but they are of little use when applied to correct
programs, since they cannot be used to prove the absence of
faults.
\end{itemize}
Here again, we argue that relative correctness can
act as a disruptive concept, since it blurs this neat separation
of duties.  In \cite{chicago2016}, we present a 
relative correctness-based static analysis 
method that enables us to locate and remove a fault from a program,
and prove that the fault has been removed ---all without testing.
This technique, which we call {\em debugging without testing},  shows that
we can apply static analysis to an incorrect program to
prove that, although it may be incorrect, it is still more correct
than another.  Given
that there are orders of magnitude more incorrect programs than
there are
correct programs, the pursuit of this idea may 
expand the scope of
static analysis methods.

In \cite{oslo2015}, we discuss how relative correctness can be used
in the derivation of a correct program from a specification.  Whereas
traditional programming calculi derive programs from specifications
by successive refinement-based correctness-preserving
transformations starting from the specification, we show that we can
derive a program by successive correctness-enhancing transformations
(using relative correctness)
starting from the trivial program {\tt abort}.  We refer to this
technique as {\em programming without refining}
\cite{oslo2015}.

In this paper, we explore the use of relative correctness in program
repair.  Specifically, we discuss how to perform program repair when
we test candidate mutants for relative correctness rather than
absolute correctness.  We are not
offering a viable, validated, empirically supported solution; rather,
we are merely analyzing current practice, discussing why we believe
a relative-correctness-based approach may offer better outcomes, and
supporting our case with analytical arguments as well
as a simple illustrative example.

In section \ref{relcorsect} we define relative correctness and
explore its main properties; since our definitions and discussions
rely on relational calculi, we devote section \ref{backgroundsect}
to a brief discussion of relational concepts.
In section \ref{repairsect} we critique the current practice of program
repair, which is based on a test of absolute correctness, and argue,
on the basis of analytical arguments,
that using a test of relative (rather than absolute)
correctness leads to better outcomes.
We complement the analytical argument of section \ref{repairsect}
by an empirical illustration in section \ref{illustrationsect} in the
form of a faulty program, which we repair in a stepwise manner
by removing its faults one by one, making it increasingly more-correct
until it becomes absolutely correct.
We summarize
and assess our findings in section \ref{conclusionsect}, and we briefly
sketch directions for future research.

\section{Relational Mathematics}
\label{backgroundsect}

We assume the reader familiar with relational algebra, and we generally
adhere to the definitions of 
\cite{schmidt90}\cite{brinkkahlschmidt97}.  Dealing with programs, 
we represent sets using a programming-like notation, by introducing variable names 
and associated data types.  If, e.g. we define set $S$ by 
the variable declarations\\
\tabeq $
	x: X; y: Y; z: Z,$\\
then $S$ is the Cartesian product $X\times Y\times Z$.  
Elements of $S$ are denoted by $s$, and are triplets of elements of  $X$, 
$Y$, and $Z$.  Given $s$ in $S$, we represent its $X$-component 
(resp. $Y$-component, $Z$-component) by (resp.) $x(s)$, 
$y(s)$, $z(s)$.  
When no risk of ambiguity exists, we may write $x$ to represent $x(s)$, and $x'$ to
represent $x(s')$.
A relation on $S$ 
is a subset of the Cartesian product $S\times S$.  Special relations on $S$ 
include the {\em universal} relation $L=S\times S$, the {\em identity} relation 
$I=\{(s,s')| s'=s\}$, and the {\em empty} relation $\emptyrelation=\{\}$.
Operations on relations (say, $R$ and $R'$) include the set theoretic operations of 
{\em union} ($R\cup R'$), {\em intersection} ($R\cap R'$),
{\em difference} ($R\setminus R'$) and {\em complement}
($\overline{R}$).  They also include the {\em relational product}, denoted by ($
R\circ R'$), or ($RR'$, for short) and defined by:  
$$RR'= \{(s,s')| \exists s'': (s,s'')\in R\wedge (s'',s')\in R'\}.$$
The {\em power} of relation $R$ is denoted by $R^n$, for a natural
number $n$, and defined by $R^0=I$, and for $n>0$, $R^n=R\circ R^{n-1}$.  The
{\em reflexive transitive closure} of relation $R$ is denoted by $R^*$ and defined
by $R^*=\{(s,s')|\exists n\geq 0: (s,s')\in R^n\}$.
The {\em converse} of relation $R$ is the relation denoted by $\widehat{R}$ and 
defined by 
$\widehat{R}=\{(s,s')| (s',s)\in R\}.$
The {\em domain} of a relation $R$ is defined as the set $\dom(R)=
\{s| \exists s': (s,s')\in R\}$, and the {\em range} of relation $R$ is 
defined as the domain of $\widehat{R}$.
Note that given a relation $R$, the product of $R$ by $L$ represents
the relation $\dom(R)\times S$; hence this expression is in effect
a relational representation of the domain of $R$; we may sometimes
use $\dom(R)$ and $RL$ interchangeably to refer to the domain of $R$.
Operator precedence is adopted as follows:  unary operators apply
first, followed by product, then intersection, then union.

A relation $R$ is said to be {\em reflexive}
if and only if $I\subseteq R$, {\em symmetric}
if and only if $R=\widehat{R}$, {\em antisymmetric} if and 
only if  $R\cap\widehat{R}\subseteq I$, 
{\em asymmetric} if and only if $R\cap\widehat{R}=\emptyrelation$, 
and {\em transitive}
if and only if $RR\subseteq R$.  A relation is said to be a {\em partial ordering}
if and only if it is reflexive, antisymmetric, and transitive.  Also, a relation $R$
is said to be {\em total} if and only if $I\subseteq R\widehat{R}$, and a relation 
$R$ is said to be {\em deterministic} (or: a {\em function}) if and only if 
$\widehat{R}R\subseteq I$.  

We use relations to represent specifications or programs.
A key concept in any study of program correctness
is the refinement ordering;  the following definition 
lays out our version
of this ordering.
\begin{definition}
\label{refinedefine}
Given two relations $R$ and $R'$, we say that
$R'$ {\em refines} $R$ (abbrev: $R'\refines R$) if and only if:
$RL\cap R'L\cap (R\cup R')=R.$
\end{definition}
Intuitively, a relation $R'$ refines a relation $R$ if and only if
it has a larger domain and assigns fewer images than $R$ to elements
of the domain of $R$.  See Figure \ref{refinefig}, where
$R''$ refines $R'$, which in turns refines $R$.

\begin{figure}
\thicklines
\setlength{\unitlength}{0.022in}
\begin{center}
\begin{picture}(170,65)

\put(0,0) {\makebox(0,0){6}}
\put(0,10){\makebox(0,0){5}}
\put(0,20){\makebox(0,0){4}}
\put(0,30){\makebox(0,0){3}}
\put(0,40){\makebox(0,0){2}}
\put(0,50){\makebox(0,0){1}}
\put(0,60){\makebox(0,0){0}}

\put(50,0) {\makebox(0,0){6}}
\put(50,10){\makebox(0,0){5}}
\put(50,20){\makebox(0,0){4}}
\put(50,30){\makebox(0,0){3}}
\put(50,40){\makebox(0,0){2}}
\put(50,50){\makebox(0,0){1}}
\put(50,60){\makebox(0,0){0}}

\put(60,0) {\makebox(0,0){6}}
\put(60,10){\makebox(0,0){5}}
\put(60,20){\makebox(0,0){4}}
\put(60,30){\makebox(0,0){3}}
\put(60,40){\makebox(0,0){2}}
\put(60,50){\makebox(0,0){1}}
\put(60,60){\makebox(0,0){0}}

\put(110,0) {\makebox(0,0){6}}
\put(110,10){\makebox(0,0){5}}
\put(110,20){\makebox(0,0){4}}
\put(110,30){\makebox(0,0){3}}
\put(110,40){\makebox(0,0){2}}
\put(110,50){\makebox(0,0){1}}
\put(110,60){\makebox(0,0){0}}

\put(120,0) {\makebox(0,0){6}}
\put(120,10){\makebox(0,0){5}}
\put(120,20){\makebox(0,0){4}}
\put(120,30){\makebox(0,0){3}}
\put(120,40){\makebox(0,0){2}}
\put(120,50){\makebox(0,0){1}}
\put(120,60){\makebox(0,0){0}}

\put(170,0) {\makebox(0,0){6}}
\put(170,10){\makebox(0,0){5}}
\put(170,20){\makebox(0,0){4}}
\put(170,30){\makebox(0,0){3}}
\put(170,40){\makebox(0,0){2}}
\put(170,50){\makebox(0,0){1}}
\put(170,60){\makebox(0,0){0}}

%  R

\multiput(5,10)(0,20){3}{\vector(4,0){40}}
\multiput(5,10)(0,20){3}{\vector(4,1){40}}
\multiput(5,10)(0,20){3}{\vector(4,-1){40}}

%  R'

\multiput(65,10)(0,10){5}{\vector(4,0){40}}
\multiput(65,10)(0,10){5}{\vector(4,1){40}}

%  R''

\multiput(125,0)(0,10){6}{\vector(4,1){40}}
\put(125,0){\vector(4,0){40}}

{\scriptsize
\put(25,65){\makebox(0,0){$R$}}
\put(85,65){\makebox(0,0){$R'$}}
\put(145,65){\makebox(0,0){$R''$}}}

\end{picture}
\end{center}
\caption{\label{refinefig}Refinement:  $R'\refines R$, $R''\refines R'$}
\end{figure}

\section{Absolute Correctness and Relative Correctness}
\label{relcorsect}

\subsection{Program Functions}
\label{programfunctionsect}

If a program $p$ manipulates variables, say $x:X$ and $y:Y$,
we say that set $S=X\times Y$ is the {\em space} of $p$ and we refer to
elements of $S$ as {\em states} of $p$.
Given a program $p$ on space $S$, we denote by $[p]$ the function that $p$
defines on its space, i.e.\\ \tabeq
$[p]=\{(s,s')| $if program $p$ executes on state $s$ then it terminates
in state $s'\}.$\\
We represent programs by means of C-like programming constructs,
which we present below along with their semantic definitions:
\begin{itemize}
\item {\em Abort:}  $[${\tt abort}$] \equiv  \emptyrelation$.
\item {\em Skip:} $[${\tt skip}$]\equiv I$.
\item {\em Assignment:} $[s=E(s)]\equiv \{(s,s')| s\in\defin(E)\wedge s'=E(s)\}$, where
$\defin(E)$ is the set of states for which expression $E$ can be evaluated.
\item {\em Sequence:} $[p_1; p_2]\equiv [p_1]\circ [p_2]$.
\item {\em Conditional:} $[\texttt{if}
~(t)~\{p\}] \equiv T\cap [p]\cup \overline{T}\cap I$, where
$T$ is the relation defined as:  $T=\{(s,s')| t(s)\}$.
\item {\em Alternation:} $[\texttt{if}
~(t)~\{p\}~else~\{q\}] \equiv
T\cap [p]\cup \overline{T}\cap[q]$, where $T$ is defined as above.
\item {\em Iteration:} $[
\texttt{while}~(t)~\{b\}]\equiv (T\cap [b])^*\cap\widehat{\overline{T}}$,
where $T$ is defined as above.
\item {\em Block:} $[\{x: X; p\}]  \equiv \{(s,s')| \exists x, x'
\in X: (\langle s,x\rangle,
\langle s',x'\rangle)\in [p]\}$.
\end{itemize}
We will usually use upper case $P$ as a shorthand for $[p]$.
By abuse of notation, we may refer to a program and its function
by the same name.

\subsection{Absolute Correctness}

\begin{definition}
\label{correctnessdef}
Let $p$ be a program on space $S$ and let $R$ be a specification on $S$.  
We
say that program $p$ is {\em correct} with respect to $R$ if and only if
$P$ refines $R$.
We say that program $p$ is {\em partially correct} with respect
to specification $R$ if and only if $P$ refines $R\cap PL$.
\end{definition}
This definition is consistent with traditional definitions of 
partial and total
correctness \cite{gries81}\cite{hehner92}\cite{dijkstra76}.  
Whenever we want to contrast correctness with partial correctness, we may
refer to it as {\em total correctness}.
The following proposition,
due to \cite{mills86}, gives a simple characterization of correctness
for deterministic programs.
\begin{proposition}
\label{correctnessprop}
\label{domrpprop}
Program $p$ is correct with respect to specification $R$ if and only if
$(R\cap P)L=RL$.
\end{proposition}
By construction, $(R\cap P)L$ is a subset of $RL$, and
correct programs are those that reach the maximum of
$(R\cap P)L$, which is $RL$.

\subsection{Relative Correctness:  Deterministic Programs}

\begin{figure*}
\thicklines
\setlength{\unitlength}{0.022in}
\begin{center}
\begin{picture}(170,65)

\put(0,0) {\makebox(0,0){6}}
\put(0,10){\makebox(0,0){5}}
\put(0,20){\makebox(0,0){4}}
\put(0,30){\makebox(0,0){3}}
\put(0,40){\makebox(0,0){2}}
\put(0,50){\makebox(0,0){1}}
\put(0,60){\makebox(0,0){0}}

\put(50,0) {\makebox(0,0){6}}
\put(50,10){\makebox(0,0){5}}
\put(50,20){\makebox(0,0){4}}
\put(50,30){\makebox(0,0){3}}
\put(50,40){\makebox(0,0){2}}
\put(50,50){\makebox(0,0){1}}
\put(50,60){\makebox(0,0){0}}

\put(60,0) {\makebox(0,0){6}}
\put(60,10){\makebox(0,0){5}}
\put(60,20){\makebox(0,0){4}}
\put(60,30){\makebox(0,0){3}}
\put(60,40){\makebox(0,0){2}}
\put(60,50){\makebox(0,0){1}}
\put(60,60){\makebox(0,0){0}}

\put(110,0) {\makebox(0,0){6}}
\put(110,10){\makebox(0,0){5}}
\put(110,20){\makebox(0,0){4}}
\put(110,30){\makebox(0,0){3}}
\put(110,40){\makebox(0,0){2}}
\put(110,50){\makebox(0,0){1}}
\put(110,60){\makebox(0,0){0}}

\put(120,0) {\makebox(0,0){6}}
\put(120,10){\makebox(0,0){5}}
\put(120,20){\makebox(0,0){4}}
\put(120,30){\makebox(0,0){3}}
\put(120,40){\makebox(0,0){2}}
\put(120,50){\makebox(0,0){1}}
\put(120,60){\makebox(0,0){0}}

\put(170,0) {\makebox(0,0){6}}
\put(170,10){\makebox(0,0){5}}
\put(170,20){\makebox(0,0){4}}
\put(170,30){\makebox(0,0){3}}
\put(170,40){\makebox(0,0){2}}
\put(170,50){\makebox(0,0){1}}
\put(170,60){\makebox(0,0){0}}

\put(5,50){\vector(4,1){40}}
\put(5,50){\vector(4,-1){40}}
\put(5,40){\vector(4,1){40}}
\put(5,40){\vector(4,-1){40}}
\put(5,30){\vector(4,1){40}}
\put(5,30){\vector(4,-1){40}}
\put(5,20){\vector(4,1){40}}
\put(5,20){\vector(4,-1){40}}
\put(5,10){\vector(4,1){40}}
%\put(5,10){\vector(4,-1){40}}

\put(125,50){\vector(4,1){40}}
\put(125,40){\vector(4,1){40}}
\put(125,30){\vector(4,1){40}}
\put(125,20){\vector(4,1){40}}
\put(125,10){\vector(4,1){40}}
\put(125,0){\vector(4,1){40}}

\put(65,60){\vector(4,-1){40}}
\put(65,50){\vector(4,-1){40}}
\put(65,40){\vector(4,-1){40}}
\put(65,30){\vector(4,-1){40}}
\put(65,20){\vector(4,-1){40}}
\put(65,10){\vector(4,-1){40}}

{\scriptsize
\put(25,65){\makebox(0,0){$R$}}
\put(85,65){\makebox(0,0){$P$}}
\put(145,65){\makebox(0,0){$P'$}}}

\put(60,35){\oval(4,40)}
\put(120,30){\oval(4,50)}

\end{picture}
\end{center}
\caption{\label{morecorrectfig}Enhancing correctness without duplicating
behavior: $P'\refines_R P$}
\end{figure*}
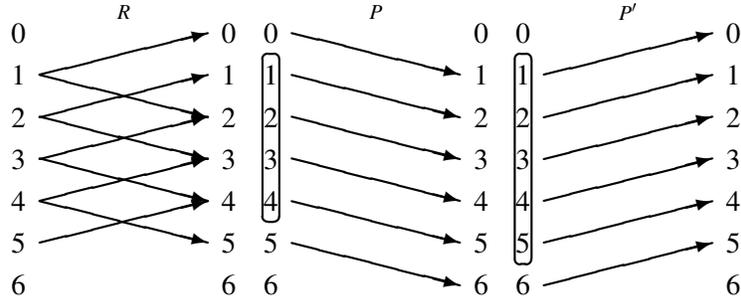

\begin{definition}
\label{relcordef}
Let $R$ be a specification on space $S$ and let $p$ and $p'$ be
two deterministic
programs on space $S$ whose functions are respectively $P$
and $P'$.  
We say that program $p'$ is {\em more-correct} than program $p$
with respect to specification $R$ (abbrev:  $P'\refines_R P$)
if and only if:
$(R\cap P')L\supseteq (R\cap P)L.$
Also, we say that program $p'$ is {\em strictly more-correct} than
program $p$
with respect to specification $R$ (abbrev:  $P'\sqsupset_R P$) 
if and only if
$(R\cap P')L\supset (R\cap P)L.$
\end{definition}
Interpretation:  $(R\cap P)L$ represents (in relational form) the set
of initial states on which the behavior of $P$ satisfies specification
$R$.  We refer to this set as the {\em competence domain} of program $P$
with respect to specification $R$.
For deterministic programs $p$ and $p'$, relative correctness of $p'$ 
over $p$ with respect to specification $R$
simply means that $p'$ has a larger competence domain than $p$.
Whenever we want to contrast correctness 
with relative correctness, we refer to it as {\em absolute correctness}.
Note that when we say {\em more-correct} we really mean {\em more-correct or
as-correct-as}.
Note also that program $p'$ may be more-correct
than program $p$ without duplicating the behavior of $p$ over
the competence domain of $p$; see Figure \ref{morecorrectfig}.
In the example
shown in this figure, we have:\\
\tabeq $(R\cap P)L = \{1,2,3,4\}\times S$,\\
\tabeq $(R\cap P')L = \{1,2,3,4,5\}\times S$,\\
where $S=\{0,1,2,3,4,5,6\}$.  
Hence $p'$ is more-correct than $p$ with respect to $R$.

\begin{figure}
\thicklines
\setlength{\unitlength}{0.012in}
\begin{center}
\begin{picture}(138,170)

\put(69,5){\line(1,1){60}}
\put(69,5){\line(-1,1){60}}
\put(69,5){\line(0,1){40}}

\put(69,125){\line(1,-1){60}}
\put(69,125){\line(-1,-1){60}}
\put(69,125){\line(0,1){40}}

\put(9,105){\line(0,-1){40}}
\put(9,105){\line(1,-1){60}}
\put(9,105){\line(1,1){60}}

\put(129,105){\line(0,-1){40}}
\put(129,105){\line(-1,-1){60}}
\put(129,105){\line(-1,1){60}}

{\scriptsize
\put(69,0){\makebox(0,0){$P_0$}}
\put(138,65){\makebox(0,0){$P_3$}}
\put(138,105){\makebox(0,0){$P_5$}}
\put(69,55){\makebox(0,0){$P_2$}}
\put(69,115){\makebox(0,0){$P_6$}}
\put(69,170){\makebox(0,0){$P_7, P_8, P_9$}}
\put(0,65){\makebox(0,0){$P_1$}}
\put(0,105){\makebox(0,0){$P_4$}}}

\put(69,170){\oval(50,17)}

\end{picture}
\end{center}
\caption{\label{treillisfig}
Ordering Candidate Programs by Relative Correctness}
\end{figure}
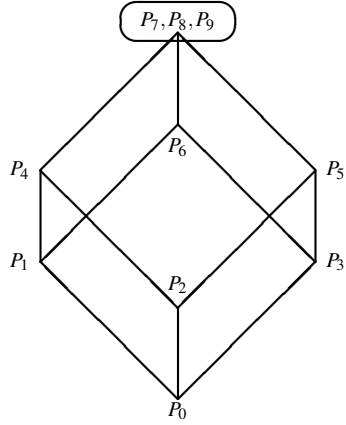

In order to highlight the contrast between relative correctness
(as a partial ordering) and absolute
correctness (as a binary attribute), we
consider the specification $R$ on space
$S=\{a,b,c,d,e\}$\\ \tabeq
$R=\{(a,a),(a,b),(a,c),(b,b),(b,c),(b,d),(c,c),(c,d),(c,e)\},$\\
and we consider the following programs, along with their
competence domains:
\begin{itemize}
\item $P_0=\{(a,d),(b,a)\}$.  $CD_0=\{\}$.
\item $P_1=\{(a,b),(b,e)\}$.  $CD_1=\{a\}$.
\item $P_2=\{(a,d),(b,c)\}$.  $CD_2=\{b\}$.
\item $P_3=\{(b,e),(c,d)\}$.  $CD_3=\{c\}$.
\item $P_4=\{(a,b),(b,c),(c,a)\}$.  $CD_4=\{a,b\}$.
\item $P_5=\{(a,d),(b,c),(c,d)\}$.  $CD_5=\{b,c\}$.
\item $P_6=\{(a,c),(b,e),(c,d)\}$.  $CD_6=\{a,c\}$.
\item $P_7=\{(a,a),(b,b),(c,c),(d,d)\}$.  $CD_7=\{a,b,c\}$.
\item $P_8=\{(a,b),(b,c),(c,d),(d,e)\}$.  $CD_8=\{a,b,c\}$.
\item $P_9=\{(a,c),(b,d),(c,e),(d,a)\}$.  $CD_9=\{a,b,c\}$.
\end{itemize}
Figure \ref{treillisfig} shows how these programs are ordered
by relative correctness with respect to $R$; in this
sample, programs $P_7, P_8, P_9$ are (absolutely) correct while programs
$P_0, P_1, P_2, P_3, P_4, P_5, P_6$ are incorrect because their competence
domain are different from (smaller than) the domain of $R$ ($\{a,b,c\}$).
See
Figure \ref{stepwisefig} for a more concrete example of programs
ordered by relative correctness.

\subsection{Relative Correctness: Non-Deterministic Programs}

We let $S$ be the space defined by non-zero natural variables $x$ and $y$ and we
let $R$ be the following specification on $R$:
$R=\{(s,s')| x'\geq 3x^2\}.$
Let $p$ and $p'$ be the following programs:\\
\tabeq {\tt p:  \{x=13*x; y=y+2*x;\}},\\
\tabeq {\tt p':  \{x=19*x; y=y+7*x;\}}.\\
The functions of these programs are:\\
\tabeq $P=\{(s,s')| x'=13\times x\wedge y'=y+26\times x\}$\\
\tabeq $P'=\{(s,s')| x'=19\times x\wedge y'=y+133\times x\}$.\\
We are interested to analyze the relative correctness of $p$ and $p'$ with
respect to $R$; to this effect, according to definition \ref{relcordef},
we must analyze the functions of $p$ and $p'$.  But the effect of these
programs on variable $y$ could not possibly be relevant to this analysis
since $R$ does not refer to $y$.  Hence it ought to be possible for us to
analyze the relative correctness of $p$ and $p'$ with respect to $R$ by
considering the effect of $p$ and $p'$ on $x$ 
alone; this yields the following
relations:\\
\tabeq $\pi=\{(s,s')| x'=13\times x\}$,\\
\tabeq $\pi'=\{(s,s')| x'=19\times x\}$.\\
Yet, we cannot apply Definition \ref{relcordef} to $\pi$ and $\pi'$
because they are not deterministic (since they fail to
specify a final value for variable $y$).  In \cite{ramics2015} we present
a definition of relative correctness that generalizes Definition
\ref{relcordef} and applies to (possibly) non-deterministic programs;
referring to the definition given in \cite{ramics2015}, we find 
(by inspecting $\pi$ and $\pi'$ rather than $P$ and $P'$) that $p'$ is
more-correct than $p$ with respect to $R$.  Indeed, the competence domain
of program $p$ with respect to $R$
(i.e. the set of initial states for which $p$ behaves
according to $R$) is characterized by the equation $13\times x\geq 3\times x^2$, which is
equivalent (since $x$ is a natural variable) to $x\leq 4$.  Likewise, we find
that the competence domain of $p'$ is characterized by the equation
$x\leq 6$.

\section{Program Repair by Relative Correctness}
\label{repairsect}

\subsection{Faults and Fault Removal}

Now that we know what it means for a program to be more correct
than another, we are in a position to define what is a fault, and
under what condition we can say that we have removed a fault.  
Any definition of a fault assumes, implicitly, some
level of granularity at which we want to define faults; a typical 
level of granularity for C-like languages is the single assignment
statement, while finer grained features include expressions or operators
within expressions.  We use the term {\em feature} to refer to any
program part, or set of program parts, at the selected level of granularity
and we present the following definition, due to \cite{ramics2014}.

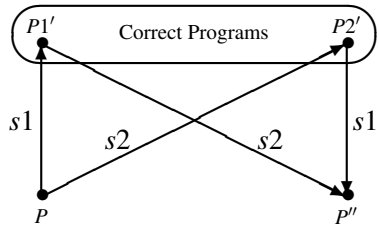
\begin{figure}
\thicklines
\setlength{\unitlength}{0.01in}
\begin{center}
\begin{picture}(200,100)

\put(17,7){\textbullet}
\put(17,87){\textbullet}
\put(177,7){\textbullet}
\put(177,87){\textbullet}

{\scriptsize
\put(20,0){\makebox(0,0){$P$}}
\put(20,100){\makebox(0,0){$P1'$}}
\put(180,0){\makebox(0,0){$P''$}}
\put(180,100){\makebox(0,0){$P2'$}}}

\put(20,10){\vector(0,1){80}}
\put(20,10){\vector(2,1){160}}
\put(180,90){\vector(0,-1){80}}
\put(20,90){\vector(2,-1){160}}

\put(10,50){\makebox(0,0){$s1$}}
\put(190,50){\makebox(0,0){$s1$}}
\put(60,40){\makebox(0,0){$s2$}}
\put(140,40){\makebox(0,0){$s2$}}

\put(100,95){\oval(190,30)}
\put(100,95){\makebox(0,0){{\scriptsize Correct Programs}}}

\end{picture}
\end{center}
\caption{\label{messyfaultsfig}
Both $P$ and $P''$: fault density of 2; fault depth of 1.}
\end{figure}

\begin{definition}
\label{faultdef}
Given a program $p$ on space $S$ and a specification $R$ on $S$,
we say that a feature $f$ of $p$ is a {\em fault} in $p$ with
respect to $R$ if and only if
there exists a feature $f'$ such that program $p'$ obtained from
program $p$ when we replace $f$ by $f'$ is strictly more-correct
than $p$ with respect to $R$.

When such an $f'$ is found, the pair $(f,f')$ is called a
{\em fault removal} in $p$ with respect to $R$.
\end{definition}
As an illustrative example, we consider the following program $p$ on space $S$
defined by variables $a$, $x$ and $i$ declared therein:
\begin{verbatim}
p:  int main () {int a[N+1]; int x=0;
            int i=0; while (i<N) {x=x+a[i]; i=i+1;}}
\end{verbatim}
and we consider the specification $R$ defined by:\\
\tabeq $R=\{(s,s')|  x'=\sum_{i=1}^N a[i]\}$.\\
The function of program $p$ is:\\
\tabeq $P=\{(s,s')|  a'=a\wedge i'=N\wedge x'=\sum_{i=0}^{N-1}a[i]\}$.\\
The competence domain of $p$ with respect to $R$ is:\\
\tabeq $(R\cap P)L=\{(s,s')| a[0]=a[N]\}$,\\
which makes sense, since this is the condition under which what
the program does (the sum of $a$ from 0 to $N-1$) coincides with what the
specification mandates (the sum from 1 to $N$).  Because the
competence domain of $p$ is distinct from $\dom(R)=S$,
this program is incorrect.  At the level of granularity of
assignment statements and logical expressions, we see two faults
in program $p$ with respect to specification $R$:
\begin{itemize}
\item The fault made up of the aggregate of statements $f1$ = {\tt (i=0, (i<N))};
the substitution  $f1'$ = {\tt (i=1, (i<N+1))} constitutes a fault removal for
$f1$.
\item The fault $f2$ = {\tt (x=x+a[i])}; substitution of $f2$ by 
$f2'$ = {\tt (x=x+a[i+1])} constitutes a fault removal for $f2$.
\end{itemize}
Note that {\tt (i=0)} alone is not a fault in $p$, nor is {\tt (i<N)} as they
admit no substitution that would make the program more-correct.  If we let
$p1'$ be the program obtained from $p$ by substituting $f1$ by $f1'$, then 
we find that $f2$ is not
a fault in $p'$, even though it is a fault in $p$; the same goes for
program $p2'$ obtained by substituting $f2'$ for $f2$.  If we
substitute $f1$ by $f1'$ and $f2$ by $f2'$ we find two faults again,
namely $f1'$ and $f2'$.
See Figure \ref{messyfaultsfig}, where we label each transformation by
the corresponding substitution, and we let $s1$ be the substitution
$(f1, f1')$ and $s2$ be the substitution $(f2, f2')$.
Even though $p$ has two faults, it is one fault removal away from
being correct; we say that it has a {\em fault density} of 2 and a 
{\em fault depth}
of 1.

\subsection{Testing for Relative Correctness}
\label{rctestingsect}

\begin{figure}
\thicklines
\setlength{\unitlength}{0.015in}
\begin{center}
\begin{picture}(200,150)

{\scriptsize
\put(100,70){\oval(160,140)}
\put(180,0){$\dom(R)$}

\put(100,70){\oval(40,80)}
\put(120,30){\makebox(0,0){{\scriptsize $CD$}}}
\put(100,70){\oval(90,120)}
\put(145,10){\makebox(0,0){{\scriptsize $CD'$}}}
\put(100,70){\oval(130,30)}
\put(165,55){\makebox(0,0){{\scriptsize $T$}}}
\put(100,70){\makebox(0,0){{\scriptsize $N_0$}}}
\put(133,70){\makebox(0,0){{\scriptsize $N_1$}}}
\put(153,70){\makebox(0,0){{\scriptsize $N_2$}}}
}
\end{picture}
\end{center}
\caption{\label{testcovfig}Test Coverage of Relative Correctness: $N=N_0+N_1+N_2$}
\end{figure}
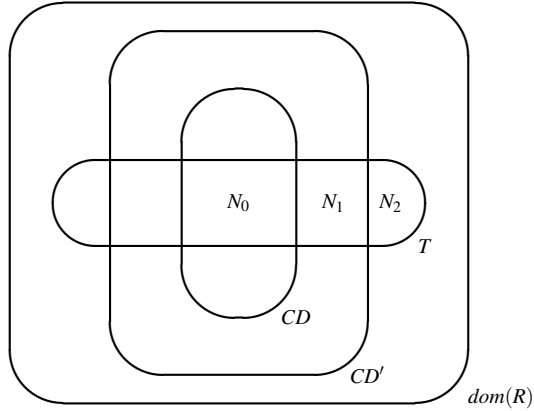

Now that we have defined what is a fault removal, we
addres the question: how do we ascertain that we have removed a fault?
As is customary for
matters pertaining to software products, we can do so in one of two
ways: 
\begin{itemize}
\item Either through a static analysis of the products' ($p$
and $p'$) source code; this is discussed in \cite{chicago2016}.
\item Or through execution and monitoring of the products in
question; this is the subject of this paper.
\end{itemize}
This raises the question:  how do we test a program $p'$ for relative
correctness over some program $p$ with respect to specification $R$,
and how is that different from testing program $p'$ for absolute
correctness with respect to $R$?
For the sake of simplicity, we address this question in the
context of deterministic programs,
and we argue that testing a program $p'$ for relative correctness over
some program $p$ with respect to some specification $R$ differs from
testing program $p'$ for absolute correctness with respect to $R$ in
three important ways:
\begin{itemize}
\item {\em Test Data Selection}.
The problem of test data selection can be formulated in the following 
generic terms:
Given a large or infinite input space, say $D$, find a representative subset
$T$ of $D$ such that analysis of the behavior of candidate programs on $T$ 
enables us to infer claims about their behavior on $D$.  Regardless of what
selection criterion is adopted to derive $T$ from $D$, testing for relative 
correctness differs from testing for absolute correctness in a fundamental
way:  for absolute correctness, the input space $D$ we are trying to approximate
is $D=\dom(R)$, whereas for relative correctness the input space 
is $D=\dom(R\cap P)$, i.e. the competence domain of $p$.
Indeed, to prove that $p'$ is more-correct than $p$ with respect to $R$,
we must prove that $p'$ runs successfully for all elements of the
competence domain $D$ of $p$, which we do by checking that $p'$ runs
successfully for all elements of $T$ (an approximation of $D$).
\item {\em Oracle Design}.
Let $\absoracle(s,s')$ be the oracle for absolute correctness derived
from specification $R$.  Then the oracle for relative correctness of
program $p'$ over program $p$, which we denote by $\reloracle(s,s')$
must ensure that program $p'$ satisfies $\absoracle(s,s')$ for all $s$
in the competence domain of $p$ with respect to $R$.  We write it as:
$$\reloracle(s,s')\equiv (\absoracle(s,P(s))\Rightarrow \absoracle(s,s')).$$
As for oracle $\absoracle(s,s')$ it must be derived from specification $R$
according to the following formula:
$$\absoracle(s,s')\equiv (s\in\dom(R)\Rightarrow (s,s')\in R).$$
Indeed, we do not want a test to fail on some input $s$ outside the
domain of $R$, as candidate programs are only responsible for their
behavior on $\dom(R)$.  Hence the condition $(s,s')\in R$ is checked
only for $s$ in $\dom(R)$; for $s$ outside the domain of $R$, the
test is (vacuously) considered successful.
\item {\em Test Coverage Assessment}.  
When we test a program $p'$ for absolute correctness with respect to
some specification $R$ using a test data
of size $N$, we gain a level of confidence in the correctness of
$p'$, to an extent that is commensurate with $N$.
On the other hand, when we test a program $p'$ for relative correctness
over a program $p$ with respect to some specification $R$ on a test data
of size $N$, $N$ does not tell the whole story:  We also need to know
whether $p'$ behaves better than $p$ because $p$ fails often or because
$p'$ succeeds often.  Hence we may need to quantify the outcome of the 
experiment by means of three variables:  $N_0$, the number of test
cases when both $p$ and $p'$ succeed; $N_1$, the number of test cases
when $p$ fails and $p'$ succeeds; and $N_2$, the number of test
cases when both fail.  While $N=N_0+N_1+N_2$ tells us to what extent
$p'$ is better than $p$ (i.e. to what extent we can be confident that
$p'$ is more-correct than $p$), the partition of $N$ into $N_0$, $N_1$
and $N_2$ tells us whether $p'$ is better than $p$ because $p'$
succeeds often or because $p$ fails often.
See Figure \ref{testcovfig}.
\end{itemize}

\subsection{Program Repair with Absolute Correctness}

Most techniques for program repair 
\cite{ArcuriYao2008}\cite{DW10}\cite{gopinath2011}\cite{KSK13}%
\cite{LeGoues+2012}\cite{Stryker} proceed by 
applying transformations on an original faulty
program. These transformations may be macro-transformations (including 
multi-site program
modifications), or micro-modifications
(intra-statement) using mutation operators such as those provided by the muJava
\cite{ma2005} program mutation tool. Two main approaches exist towards assessing
the suitability of the generated transformations: test-based techniques
\cite{ArcuriYao2008}\cite{DW10}\cite{KSK13}\cite{LeGoues+2012} 
(which use the successful
execution of the candidate program on a test suite
as the acceptance criterion) or specification-based techniques 
\cite{gopinath2011}\cite{Stryker}
(which use a specification and some sort of 
constraint-solving to determine if the new code complies with the 
specification).  In both cases, mutants are selected on the basis of
an analysis of their absolute correctness with respect to the
specification at hand (embodied in the oracle in the case of testing).

\begin{figure}
\thicklines
\setlength{\unitlength}{0.015in}
\begin{center}
\begin{picture}(260,130)

{\scriptsize
\put(50,70){\oval(60,30)}
\put(80,55){\makebox(0,0){{\scriptsize $T$}}}
\put(50,70){\oval(100,40)}
\put(100,50){\makebox(0,0){{\scriptsize $CD'$}}}
\put(50,70){\oval(40,80)}
\put(70,30){\makebox(0,0){{\scriptsize $CD$}}}
\put(50,70){\oval(120,120)}
\put(200,70){\oval(120,120)}

\put(200,70){\oval(30,60)}
\put(215,40){\makebox(0,0){{\scriptsize $CD$}}}
\put(200,70){\oval(50,100)}
\put(225,20){\makebox(0,0){{\scriptsize $CD'$}}}
\put(200,70){\oval(90,30)}
\put(245,55){\makebox(0,0){{\scriptsize $T$}}}

\put(100,8){$\dom(R)$}
\put(250,8){$\dom(R)$}
}

\put(50,2){{\scriptsize (a)}}
\put(200,2){{\scriptsize (b)}}
\end{picture}
\end{center}
\caption{\label{wrongmutantfig}Absolute Correctness-based
Repair:  Poor Precision, Poor Recall}
\end{figure}
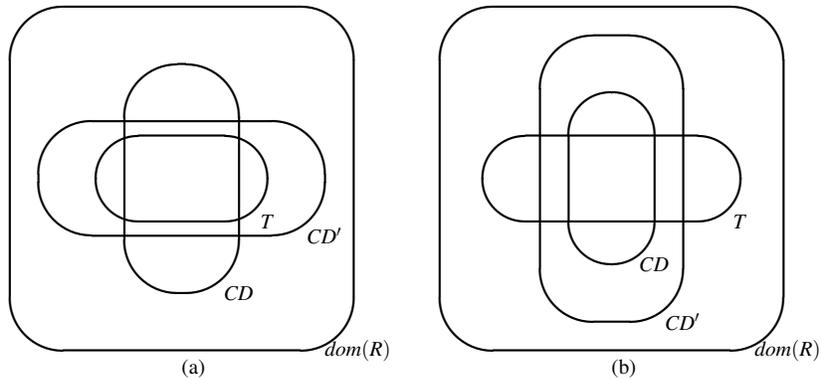

We argue that selecting mutants on the basis of their absolute correctness
is flawed, because when we remove a fault from the original program, we
have no reason to believe that the new program is correct, unless we
assume that the fault that we have just removed is the program's last
fault.  Instead, the best we can hope for when we generate a mutant from
a base program is that the mutant is more-correct than the base program
with respect to the specification at hand; consequently, we should be
testing mutants for relative correctness rather than absolute correctness.
Specifically, we argue that
when mutants are evaluated on the basis of their 
absolute correctness on a sample
test data $T$, both the decision to retain successful mutants and the 
decision to reject unsuccessful mutants, are wrong:
\begin{itemize}
\item As Figure \ref{wrongmutantfig}(a) shows (if $\mathit{CD}$ is the competence domain
of the original program and $\mathit{CD}'$ is the competence domain of the mutant), a mutant
may pass the test $T$ (since $T\subseteq \mathit{CD}'$)
yet not be
more-correct than the original (since $\mathit{CD}$ is not a subset of $\mathit{CD}'$).
\item As Figure \ref{wrongmutantfig} (b) shows, a mutant may fail the test $T$
(since $T$ is not a subset of $\mathit{CD}'$) and
yet still be more-correct than the original (since $\mathit{CD}\subseteq \mathit{CD}'$).
\end{itemize}
As a result, neither the precision nor the recall of the selection algorithm is
assured.

\subsection{Program Repair with Relative Correctness}

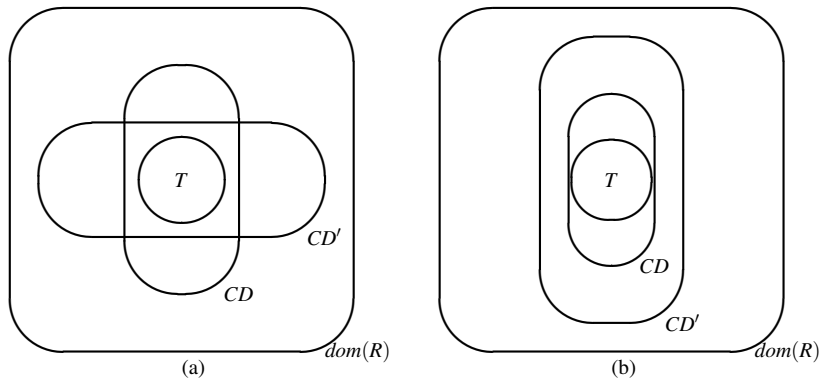
\begin{figure}
\thicklines
\setlength{\unitlength}{0.015in}
\begin{center}
\begin{picture}(260,130)

{\scriptsize
\put(50,70){\oval(30,30)}
\put(50,70){\makebox(0,0){{\scriptsize $T$}}}
\put(50,70){\oval(100,40)}
\put(100,50){\makebox(0,0){{\scriptsize $CD'$}}}
\put(50,70){\oval(40,80)}
\put(70,30){\makebox(0,0){{\scriptsize $CD$}}}
\put(50,70){\oval(120,120)}
\put(200,70){\oval(120,120)}

\put(200,70){\oval(30,60)}
\put(215,40){\makebox(0,0){{\scriptsize $CD$}}}
\put(200,70){\oval(50,100)}
\put(225,20){\makebox(0,0){{\scriptsize $CD'$}}}
\put(200,70){\oval(28,28)}
\put(200,70){\makebox(0,0){{\scriptsize $T$}}}

\put(100,8){$\dom(R)$}
\put(250,8){$\dom(R)$}
}

\put(50,2){{\scriptsize (a)}}
\put(200,2){{\scriptsize (b)}}
\end{picture}
\end{center}
\caption{\label{rightmutantsfig}Relative Correctness-based
Repair:  Poor Precision, Perfect Recall}
\end{figure}

In light of the foregoing discussion, we argue that if a fault removal
is expected to
make a program more-correct than the original (vs absolutely correct)
then it is only fair that it be tested for relative correctness rather
than absolute correctness.  In this way, if a program has several faults,
we can remove them one at a time in a stepwise manner.  To do so, we
adopt the test data selection criterion and the oracle design discussed
in section \ref{rctestingsect}.
As we recall, in order to test a program
for relative correctness over a program $P$ with respect to a specification
$R$, we have to select test data in the competence domain of $P$ with
respect to $R$ ($CD$ in Figure \ref{rightmutantsfig}) rather
than to select it in $\dom(R)$.  As we can see from Figure
\ref{rightmutantsfig}(b), this ensures perfect recall since all programs
that are relativeley correct with respect to $P$ are selected.  As for
retrieval precision, it depends on the quality of the test data, but
as Figure \ref{rightmutantsfig}(a) shows, we can still select programs
that are not more-correct than $P$, if $T$ is not adequately distributed
over $CD$; hence precisions remains an issue.

\section{Illustration:  Fermat Decomposition}
\label{illustrationsect}

\subsection{Experimental Setup}

To illustrate the distinction between program repair by absolute
correctness and by relative correctness, we consider a program
that performs the Fermat decomposition of a natural number, in
which we introduce three changes.
The space of a Fermat decomposition is defined
by three natural variables, $n$, $x$ and $y$ and the specification
is defined as follows:
$$R=\{(s,s')| ((n \modk 2=1)\lor (n \modk 4=0)) \wedge n=x'^2-y'^2\}.$$
A correct Fermat program (which we call $p'$) is:
\begin{verbatim}
void fermatFactorization() { 
   int n, x, y; // input/output variables
   int r; // work variable
   x = 0; r = 0;
   while (r < n) {r = r + 2 * x + 1; x = x + 1; }
   while (r > n) {int rsave; y = 0; rsave = r;
      while (r > n) {r = r - 2 * y - 1; y = y + 1; }
      if (r < n) {r = rsave + 2 * x + 1; x = x + 1; }}}
\end{verbatim}
The three changes we introduce in this program are shown below;
we do not call them faults yet because we do not know whether they
meet our definition of a fault (Definition \ref{faultdef}).
A given number of changes (re: three in this case) can lead to
fewer faults (if some changes cancel each other, or if one or
more changes have no effect on the function of the program);
also, a given number of changes (three in this case) can also
lead to a larger number of faults (the same change can be
remedied either by reversing the change or by altering the
program elsewhere to cancel the change).  We revisit this discussion
in the next section.

We let $p$ be the program obtained after introducing the changes to $p'$:
\begin{verbatim}
void basep(int& n, int& x, int& y) {
   int r; x = 0; r = 0;
   while (r < n) {r = r + 2 * x - 1; /*change in r*/ x =x+1;}
      while (r > n) {int rsave; rsave = r; y = 0;
         while (r > n) {r =r-2*y+1; /*change in r*/ y =y+1;}
         if (r < n) {r =rsave+2*x-1; /*change in r*/ x =x+1;}}}
\end{verbatim}
Most program repair methods proceed by generating mutants of the base
program and testing them for absolute correctness;  all we are
advocating in this paper is that instead of testing mutants for
absolute correctness, we ought to test them for relative correctness.
To illustrate our approach, we generate mutants of program $p$,
test them for absolte correctness, and show that none of them are
(absolutely) correct.  If absolute correctness were our only criterion,
then this would be the (unsuccessful) end of the experiment.
But we find that while none of the mutants are absolutely correct,
some are strictly more-correct than $p$; hence the transition from $p$
to these mutants represents a fault removal (by Definition \ref{faultdef}).
If we take these mutants as our base programs and apply the mutation
generator to them, then test them for strict relative correctness,
we can iteratively remove the faults of the
program in a stepwise manner, climbing the relative correctness
ordering until we reach a (absolutely) correct program.  

Specifically, we start from program $p$
and apply muJava to generate mutants using
the single mutation option with the AORB operator (Arithmetic 
Operator Replacement, Binary).  Whenever a set of mutants are
generated, we subject them to three tests:  
\begin{itemize}
\item A test for absolute correctness, using the oracle
$\absoracle(s,s')$.
\item A test for relative correctness, using the oracle
$\reloracle(s,s')$.
\item A test for strict relative correctness, which in addition
to relative correctness checks the presence of at least one state
in the competence domain of the mutant that is not in the
competence domain of the base program.
\end{itemize}
The mutants that are found to be strictly more-correct than the
base program are used as new base programs, and the process is
iterated again until at least one mutant is found to be absolutely
correct; we select this mutant as the repaired version of the original
program $p$.
The main iteration of the test driver is given below.
All the details of our experiment are
posted online at \url{https://selab.njit.edu/programrepair/}.
\begin{verbatim}
int main ()
 {for (int mutant =1; mutant<= nbmutants; mutant++)
  {// test mutant vs spec. R for abs and rel correctness
   bool cumulabs=true; bool cumulrel=true; bool cumulstrict=false;
   while (moretestdata) 
    {int n,x,y; int initn,initx,inity; //initial, final states
     bool abscor, relcor, strict;
     initn=td[tdi]; tdi++;  // getting test data
     n=initn; x=initx; y=inity;   // saving initial state
     callmutant(mutant, n, x, y);
     abscor = absoracle(initn, initx, inity, n, x, y);
     cumulabs = cumulabs && abscor;
     n=initn; x=initx; y=inity;  //  re-initializing
     basep(n, x, y);
     relcor = ! absoracle(initn, initx, inity, n, x, y) || abscor;
     strict = ! absoracle(initn, initx, inity, n, x, y) && abscor;
     cumulrel = cumulrel && relcor;
     cumulstrict = cumulstrict || strict;
 }}}
bool R (int initn, int initx, int inity, int n, int x, int y)
   {return ((initn%2==1) || (initn%4==0)) && (initn==x*x-y*y);}
bool domR (int initn, int initx, int inity)
   {return ((initn%2==1) || (initn%4==0));}
bool absoracle (int initn,int initx,int inity,int n,int x,int y)
   {return  (! (domR(initn, initx, inity)) 
             || R(initn, initx, inity, n, x, y));}
\end{verbatim}
The main program includes two nested loops; the outer loop iterates over
mutants and the inner loop iterates over test data.  For each mutant
and test datum, we 
execute the mutant and the base program on the test datum and test the
mutant for absolute correctness ({\tt abscor}), relative correctness
({\tt relcor}) and strict relative correctness ({\tt strict}); these
boolean results are cumulated for each mutant in variables {\tt cumulabs},
{\tt cumulrel} and {\tt cumulstrict}, and are used to diagnose the mutant.
As for the Boolean functions {\tt R}, {\tt domR} and {\tt absoracle}, they
stem readily from the definition of $R$ and from the
oracle definitions given in section \ref{rctestingsect}.

\subsection{Experimental Results}

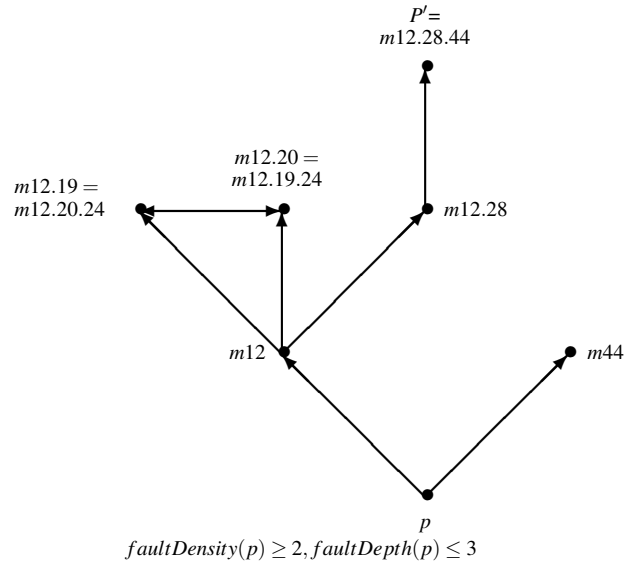
\begin{figure}
\thicklines
\setlength{\unitlength}{0.025in}
\begin{center}
\begin{picture}(130,110)

{\scriptsize
\put(91,5){\makebox(0,0){$p$}}
\put(91,110){\makebox(0,0){\shortstack{$P'$=\\$m12.28.44$}}}
\put(125,40){$m44$}
\put(50,40){$m12$}
\put(5,70){\shortstack[l]{$m12.19=$\\$m12.20.24$}}
\put(95,70){$m12.28$}
\put(60,80){\makebox(0,0){\shortstack{$m12.20=$\\$m12.19.24$}}}
}
\put(90,10){\textbullet}
\put(60,40){\textbullet}
\put(120,40){\textbullet}
\put(30,70){\textbullet}
\put(60,70){\textbullet}
\put(90,70){\textbullet}
\put(90,100){\textbullet}

\put(91,11){\vector(1,1){30}}
\put(91,11){\vector(-1,1){30}}
\put(61,41){\vector(1,1){30}}
\put(61,41){\vector(0,1){30}}
\put(61,41){\vector(-1,1){30}}
\put(91,71){\vector(0,1){30}}
\put(31,71){\vector(1,0){30}}
\put(61,71){\vector(-1,0){30}}

\put(65,0){\makebox(0,0){{\scriptsize $faultDensity(p)\geq 2,
faultDepth(p)\leq 3$}}}

\end{picture}
\end{center}
\caption{\label{stepwisefig}Relative Correctness-based
Repair:  Stepwise Fault Removal}
\end{figure}

Starting with program $p$, we apply muJava repeatedly to generate mutants,
taking mutants which are found to be strictly more-correct as base programs
and repeating
until we generate a correct program.  This proceeds
as follows:
\begin{itemize}
\item
When muJava is executed on program $p$, it produces 48 mutants, of which
two ($m12$ and $m44$)
are found to be strictly more-correct than $p$, and none are found
to be absolutely correct with respect to $R$; we pursue the analysis of
$m12$ and $m44$.
\item {\em Analysis of $m44$}.  When we apply muJava to $m44$, we find
48 mutants, none of them prove to be absolutely correct, nor
relatively correct, nor strictly relatively correct.
\item {\em Analysis of $m12$}.  We find by inspection that $m12$
reverses one of the modifications we had applied to $p'$ to find
$p$; since $m12$ is strictly more-correct than $p$ with respect to
$R$, we conclude that the feature in question was in fact a fault in
$p$ with respect to $R$.  When we apply muJava to $m12$, it generates
48 mutants, three of which prove to be strictly more-correct than
$m12$:  we name them $m12.19$, $m12.20$ and $m12.28$.  
All the other mutants are found to be neither absolutely correct
with respect to $R$, nor more correct than $m12$.
\begin{itemize}
\item {\em Analysis of $m12.19$}.  When we apply muJava to $m12.19$, it
generates 48 mutants, none of which is found to be absolutely correct
nor strictly more-correct than $m12.19$, but one ($m12.19.24$) proves
to be identical to $m12.20$ and is more-correct than (but not
strictly more-correct than, hence as correct as) $m12.19$.
\item {\em Analysis of $m12.20$}.  When we apply muJava to $m12.20$, it
generates 48 mutants, none of which is found to be absolutely correct
nor strictly more-correct than $m12.20$, but one ($m12.20.24$) proves
to be identical to $m12.19$ and is more-correct than (but not
strictly more-correct than, hence as correct as) $m12.20$.
\item {\em Analysis of $m12.28$}.  We find by inspection that $m12.28$
reverses a second modification we had applied to $p'$ to obtain $p$;
since $m12.28$ is strictly more-correct than $m12$, this feature is
a fault in $m12$; whether it is a fault in $p$ we have not checked,
as we have not compared $m12.28$ and $p$ for relative correctness.
When we apply muJava to $m12.28$, we find a single mutant, namely
$m12.28.44$ that is absolutely correct with respect to $R$, 
more-correct than $m12.28$ with respect to $R$, and strictly more-correct
than $m12.28$ with respect to $R$.
\begin{itemize}
\item {\em Analysis of $m12.28.44$}.  We find by inspection that
$m12.28.44$ is nothing but the original Fermat decomposition program
we have started out with:  $p'$.  
\end{itemize}
\end{itemize}
\end{itemize}
The results of this analysis are represented in Figure 
\ref{stepwisefig}.
Note that $m12$ and $m44$ are strictly more-correct than $p$ with respect to
$R$; hence (according to Definition \ref{faultdef}) the mutations that
produced these programs from $p$ constitute fault removals;  whence we
can say that $p$ has at least two faults, which we write as
$faultDensity(p)\geq 2$.  On the other hand, this experiment shows
that we can generate a correct program ($p'$) from $p$ by means of
three fault removals; if we let the {\em Fault Depth} of a program
be the minimal number of fault removals that separate it from a 
correct program, then we can write:  $faultDepth(p)\leq 3$.

\section{Conclusion}
\label{conclusionsect}

In this paper we discuss how we can use the concept of relative correctness
to refine the technique of program repair by mutation testing.  We argue that
when we remove a fault from a program, in the context of program repair, we
have no reason to expect the resulting program 
to be correct unless we know (how do we ever?)
that the fault we have just removed is the last fault of the program.  
Therefore we should, instead, be testing the program for relative correctness
rather than absolute correctness.  We have found that testing a program
for relative correctness rather than absolute correctness has an impact on
test data selection as well as oracle design, and have discussed practical
measures to this effect.  As an illustration of our thesis, we take a
simple example of a faulty program, which we can repair in a stepwise
manner by seeking to derive successively more-correct mutants; by contrast,
the test for absolute correctness keeps excluding all the mutants
except the last, and fails to recognize that some mutants, while
being incorrect, are still increasingly more correct than the original.
We are not offering a seamless validated solution as much as we are
seeking to draw attention to some opportunities for enhancing the
practice of software testing.

Our research agenda includes further exploration of the technique proposed
in this paper to assess its feasibility and effectiveness on software
benchmarks, as well as techniques to streamline test data selection to
enhance the precision of relative-correctness-based program repair
(re:  Figure \ref{rightmutantsfig}).

Other researchers \cite{logozzo2014}\cite{lahiri2013}\cite{logozzo2012}
have introduced a concept of relative correctness and have
explored this concept in the context of program repair.  Our
work differs significantly from theirs
in many ways:  we represent specifications by relations whereas
they specify them with assertions; we capture program semantics with
input/output functions whereas they capture them by means of
execution traces; we define relative correctness by means of
competence domains and specification violations
whereas they define it by means of correct
traces and incorrect traces; we introduce relative correctness as a way to
define faults whereas they introduce it as a way to compare program versions;
we have explored the implications of relative correctness on several 
aspects of software engineering, whereas they focus primarily on
sooftware testing.

This paper complements our earlier work in the following manner:  In
\cite{ramics2014} we introduce relative correctness for deterministic
programs, and explore the mathematical properties of this concept;
in \cite{ramics2015} we generalize the concept of relative correctness
to non-deterministic programs and study its mathematical properties.
In \cite{oslo2015} ({\em Programming without Refinement})
we argue that while we generally think of program
derivation as the process correctness preserving transformations
using refinement, it is
possible to derive programs by correctness-enhancing transformations
using relative correctness; one of the interesting advantages of
relative correctness-based correctness enhancing transformations is
that they capture, not only the derivation of programs from scratch,
but also virtually all software maintenance activities.  We can
argue in fact that software evolution and maintenance is nothing but
an attempt to enhance the correctness of a software product with 
respect to a specification.  
In \cite{chicago2016} ({\em Debugging without Testing})
we show how relative correctness can be used to
define faults and fault removals, and that we can use these definitions
to remove a fault from a program and prove that the fault has ben removed,
all by static analysis, without testing.  

\subsection*{Acknowledgements}

The authors are very grateful to the anonymous reviewers for their
insightful feedback, which has greatly contributed to the clarity
and content of this paper.

%\input{txt}
%\bibliography{../../../ref/ref}
%\bibliographystyle{eptcs}
\end{document}